# Recommender Systems Notation

Proposed Common Notation for Teaching and Research

*Michael D. Ekstrand[1] and Joseph A. Konstan[2]*

**INTRODUCTION**

As the field of recommender systems has developed, authors have used a myriad of notations for describing the mathematical workings of recommendation algorithms. These notations appear in research papers, books, lecture notes, blog posts, and software documentation. The disciplinary diversity of the field has not contributed to consistency in notation; scholars whose home base is in information retrieval have different habits and expectations than those in machine learning or human-computer interaction.

In the course of years of teaching and research on recommender systems, we have seen the value in adopting a consistent notation across our work. This has been particularly highlighted in our development of the *Recommender Systems* MOOC on Coursera (Konstan et al. 2015), as we need to explain a wide variety of algorithms and our learners are not well-served by changing notation between algorithms.

In this paper, we describe the notation we have adopted in our work, along with its justification and some discussion of considered alternatives. We present this in hope that it will be useful to others writing and teaching about recommender systems. This notation has served us well for some time now, in research, online education, and traditional classroom instruction. We feel it is ready for broad use.

**DESIGN GOALS**

We have several design goals in order to support a wide range of recommender exposition. Some of these are in conflict, and we navigate tensions between competing desiderata in some of our specific decisions.

**Flexibility**

We desire our notation to be flexible to a wide range of algorithms; it should apply equally well to neighborhood models and advanced learning-to-rank applications.

**Clarity**

We want to minimize the guesswork required in order to interpret notation. When feasible, common symbols should be connected to their referents. Notation should not be ambiguous.

---

[1] People and Information Research Team, Boise State University. michaelekstrand@boisestate.edu
[2] GroupLens Center for Social and Human-Centered Computing, University of Minnesota. konstan@umn.edu

**Concision**

At the same time, notation should not be overly verbose. While we favor clarity over terseness, clarity is not always well-served by cumbersome explicit notation.

**Commonality**

When practical, we also seek to use common notation from other fields. For example, if it is feasible to reuse common linear algebra notation, we seek to do so.

**Usability by Hand**

To support educational use, as well as whiteboard collaboration in office and lab environments, we want our notation to be legible when hand-written. In particular, we avoid distinctions in symbols that are difficult to replicate in handwriting.

**RECOMMENDATION INPUTS**

With these in mind, we begin with notation for the underlying objects in a recommender system.

We denote **items** $i, j \in I$ and **users** $u, v \in U$. We consistently maintain a distinction between the variables used to denote users and items to avoid ambiguity. If we need more than two users or items, we use numeric subscripts, such as $i_1, i_2, \ldots, i_n$.

User-item preference data is denoted in the form of a $|U| \times |I|$ **ratings matrix** $R$. $R$ is sparsely observed; in a mild abuse of notation, we also consider $R$ as a set, and write $r_{ui} \in R$ to denote that we have a rating or other observed or inferred preference from user $u$ for item $i$.

We can then denote subsets of these sets. $R_u$ is the set of ratings by user $u$ and $R_i$ is the set of ratings of item $i$. Since we use distinct variables for user and items, the direction of the subset operation is clear from context. We can also denote user and item rating vectors by $\vec{r}_u$ and $\vec{r}_i$, which again are sparsely observed. In typed material we will sometimes use boldface $\mathbf{r}_u$ and $\mathbf{r}_i$.

We find it useful to also denote subsets of users and items. $I_u = \{i \in I : r_{ui} \in R_u\}$ is the set of items that have been rated by $u$, and likewise $U_i = \{u \in U : r_{ui} \in R_i\}$ is the set of users who have rated $i$.

**Overload Note:** Our use of $I$ for the set of items conflicts with the common use of $I$ as the identity matrix. We find item sets to be a significantly more common notational need than identity matrices. Therefore, when an identity matrix is required, the less common but precedented notation of $E$ or $\mathbf{1}$ may be used.

**Summary Statistics**

This notation lends itself well to a number of summary statistics:

$|I_u|$       The number of items rated or consumed by user $u$.

| | |
|---|---|
| $\|U_i\|$ | The number of users who have rated or consumed item $i$. |
| $\|R_u\|$ | The number of ratings provided by $u$ ($\|R_u\| = \|I_u\|$ unless there are repeated ratings for the same user-item pair). |
| $\|R_i\|$ | The number of ratings provided for $i$ (likewise, $\|R_i\| = \|U_i\|$ in the absence of repeated ratings) |
| $\bar{r}_u, \bar{r}_i$ | The average of user $u$ or item $i$'s ratings: $\bar{r}_u = \frac{\sum_{i \in I_u} r_{ui}}{\|R_u\|}$ |

**Unobserved Underlying Data**

It is sometimes useful to refer to underlying "true preference" scores of which ratings are a noisy observation. We denote these scores $\pi_{ui} \in \Pi$. If we want to model a rating as being preference plus Gaussian noise (psychologically unrealistic but precedented in the recommender systems literature), we can say:

$$r_{ui} = \pi_{ui} + \epsilon_{ui}$$
$$\epsilon_{ui} \sim \text{Normal}(0, \sigma)$$

**RECOMMENDATION OUTPUTS**

Many recommendation algorithms compute an *ordering* of items for a user. The recommendation request may also be associated with a context and/or an explicit representation of a user information need, such as a search query. We can denote this with an ordering function $O$:

$$O(i|u, h, x): I \times U \times H \times X \to I^*$$

Where $H$ is a set of queries or task descriptions, and $X$ a set of contexts or context cues.

Recommendation is often performed by a top-N ranking using some per-item score; such a score is also the basis of rating prediction or preference estimation. We can similarly denote this with a function $s$:

$$s(i|u, h, x): I \times U \times H \times X \to \mathbb{R}$$

For both of these functions, any given implementation may only depend on a subset of the input variables. This enables our notation to encompass a wide range of specific applications in the recommendation and search space; for example:

- $s(i|u)$ — a traditional personalized recommender
- $s(i|h)$ — a traditional search engine
- $s(i|u, x)$ — a context-aware search engine
- $s(i|u, h, x)$ — context-aware personalized search

**Note:** we have considered several different variables to denote the set of queries or task descriptions. Earlier versions of this notation used $Q$, but that overloads with the common use of $Q$ as the right-hand side of a matrix decomposition. We also considered $T$ for task, but that would result in denoting individual task descriptions by $t$, which causes unfortunate overload with the near-universal use of $t$ for time when conducting temporal analysis, evaluations, or algorithm implementations. We choose $H$ as being a relatively neutral selection that doesn't conflict with other common use cases.

## NOTATING ALGORITHM FAMILIES

With this general notation in place, we can now apply it to describing various standard recommendation algorithms.

### Bias Model

Many algorithms build on a *user-item bias model*, or *personalized mean*; it is a useful fallback for predicting ratings when a more sophisticated algorithm cannot produce a score, and it is useful in normalization steps prior to running other algorithms (M. D. Ekstrand, Riedl, and Konstan 2010; Funk 2006). We notate this as:

$$b_{ui} = b + b_i + b_u$$
$$b = \bar{r} = \frac{\sum_{r_{ui} \in R} r_{ui}}{|R|}$$
$$b_i = \frac{\sum_{u \in U_i}(r_{ui} - b)}{|R_i| + \alpha_i}$$
$$b_u = \frac{\sum_{i \in I_u}(r_{ui} - b_i - b)}{|R_u| + \alpha_u}$$

The regularizing constants $\alpha_i$ and $\alpha_u$ determine the bias model's skepticism towards extreme biases on low-information users and items. The bias model can also be determined by optimization instead of the direct formulas above, either on its own or as a part of the optimization of some larger scoring model.

### Probabilistic Models

We can similarly denote probabilistic models using the probability that an item is in a rating set, as in this non-personalized popularity model (probability taken over users $u \in U$):

$$s(i) = \Pr[i \in I_u]$$

We can likewise write association rule metrics such as lift (in this formula, the context $h$ is the item $j$ for which lift is being used to compute related items):

$$s(i|j) = \frac{\Pr[i \in I_u | j \in I_u]}{\Pr[i \in I_u]}$$

When clear in the surrounding context, we sometimes simplify to write $\Pr[i|j]$.

**Neighborhood Approaches**

User-based nearest-neighbor scoring (Herlocker, Konstan, and Riedl 2002) can be notated as:

$$s(i|u) = \bar{r}_u + \frac{\sum_{v \in N(u|i)} w_{uv}(r_{vi} - \bar{r}_v)}{\sum_{v \in N(u|i)} |w_{uv}|}$$

This introduces two more pieces of notation. $w_{uv}$ is the interpolation weight between users $u$ and $v$; we prefer this notation over a similarity notation $s_{uv}$ or $s(u,v)$ so that we can use $s$ to denote a score and to facilitate the use of other interpolation weighting schemes without changing the overall notation. A good choice is the cosine of normalized rating vectors (M. D. Ekstrand et al. 2011), described for item weights below; many implementations use the Pearson correlation (Herlocker, Konstan, and Riedl 2002).

$N(u|i)$ is the neighborhood for $u$ with respect to $i$. This will typically be the $k$ users most similar to $u$ that have rated $i$. We can also consider the decontextualized neighborhood $N(u)$, if we simply want to find similar users but do not need them to have rated any particular items. In cases where it is useful to explicitly notate the neighborhood size, it can be done with a subscript, as in $N_k(u|i)$.

Item-based nearest-neighbor (Sarwar et al. 2001) can be described as:

$$s(i|u) = \bar{r}_i + \frac{\sum_{j \in N(i|u)} w_{ij}(r_{uj} - \bar{r}_j)}{\sum_{j \in N(i|u)} |w_{ij}|}$$

$$w_{ij} = \frac{\vec{\tilde{r}}_i \cdot \vec{\tilde{r}}_j}{\|\vec{\tilde{r}}_i\|_2 \|\vec{\tilde{r}}_j\|_2}$$

$$= \frac{\sum_{u \in U_i \cap U_j} \tilde{r}_{ui} \tilde{r}_{uj}}{\sqrt{\sum_{u \in U_i} \tilde{r}_{ui}^2} \sqrt{\sum_{u \in U_j} \tilde{r}_{uj}^2}}$$

$\vec{\tilde{r}}_i$ here denotes a normalized version of rating vector $\vec{r}_i$; often $\tilde{r}_{ui} = r_{ui} - \bar{r}_i$. The weights and neighborhoods here are analogous to those in the user-based case. $w_{ij}$ is the weight between items $i$ and $j$; while it is often computed as the cosine above, SLIM (Ning and Karypis 2011) provides an alternative using elastic net regression to learn each item's neighbor weights.

$N(i|u)$ is the neighborhood for $i$ with respect to $u$; this will usually be a subset of a larger pool of neighbors $N(i)$. Both sets are typically the items most similar to $i$ (that have been rated by $u$, in the $N(i|u)$ case), but other methods such as SLIM's lasso regularization are precedented.

**Matrix Factorization**

Matrix factorization (Koren, Bell, and Volinsky 2009) typically breaks down the rating matrix into composite user-feature and item-feature matrices:

$$R \approx PQ^\mathrm{T}$$

In this decomposition, $P$ is the $|U| \times k$ user-feature matrix and $Q$ is the $|I| \times k$ item-feature matrix. We use the $PQ^\mathrm{T}$ framing so that users and items are both represented by row vectors for consistency[3]. Within these matrices, we use standard matrix entry notation to denote subsets of this matrix:

| | |
|---|---|
| $\vec{p}_u$ or $\mathbf{p}_u$ | User $u$'s latent feature vector. |
| $\vec{q}_i$ or $\mathbf{q}_i$ | Item $i$'s latent feature vector. |
| $p_{uf}$ | User $u$'s value for feature $f$. |
| $q_{if}$ | Item $i$'s value for feature $f$. |

We can then denote a score:

$$s(i|u) = \vec{p}_u \cdot \vec{q}_i$$

We prefer dot product notation over the matrix multiplication equivalent ($\mathbf{p}_u \mathbf{q}_i^\mathrm{T}$, since latent feature vectors are row vectors) because it is easier for students without deep intuitive fluency in linear algebra to interpret — it is syntax sugar for a simple sum, rather than a special case of a more advanced sum.

This notation works well for a variety of matrix factorization applications, including those for both explicit and implicit feedback, and for matrix factorization internals for more sophisticated algorithms such as BPR-MF (Rendle et al. 2009).

**NEXT STEPS**

We hope that the recommender systems research and education community finds this useful. We do not intend to propose this as a formal standard for the community, but we have found it helpful to standardize notation across our own work and think there is merit in greater commonality in notation across the field at large.

---

[3] Since most matrix factorization techniques used in practice do not maintain feature weights in a separate singular value matrix, we omit it. When using an explicitly weighted model such as a true singular value decomposition, we can write $R \approx P \Sigma Q^T$ where $\Sigma \in \mathbb{R}^{k \times k}$ is the diagonal matrix of singular values.